\documentclass[a4paper]{article}
\usepackage{url}
\usepackage{ISCSLP2024}
\usepackage{CJKutf8}
\title{ICAGC 2024: Inspirational and Convincing Audio Generation Challenge 2024}
\name{Ruibo Fu$^{1,\ast}$\thanks{*Corresponding author}, Rui Liu$^2$, Chunyu Qiang$^3$, Yingming Gao$^4$, Yi Lu$^{1,6}$, Shuchen Shi$^7$, Tao Wang$^1$, Ya Li$^4$, Zhengqi Wen$^1$, Chen Zhang$^3$, Hui Bu$^5$, Yukun Liu$^6$, Xin Qi$^{1,6}$, Guanjun Li$^1$}
\address{
  $^1$Institute of Automation, Chinese Academy of Sciences  
  $^2$Inner Mongolia University \\
  $^3$Kuaishou Technology Co., Ltd 
  $^4$Beijing University of Posts and Telecommunications  \\
  $^5$AIShell Inc.
  $^6$University of Chinese Academy of Sciences
  $^7$Shanghai Polytechnic University }
\email{ruibo.fu@nlpr.ia.ac.cn, liurui\_imu@163.com, qiangchunyu@kuaishou.com, yingming.gao@bupt.edu.cn, yli01@bupt.edu.cn, zqwen@nlpr.ia.ac.cn}
\begin{document}

\maketitle
\begin{abstract}
  The Inspirational and Convincing Audio Generation Challenge 2024 (ICAGC 2024) is part of the ISCSLP 2024 Competitions and Challenges track. While current text-to-speech (TTS) technology can generate high-quality audio, its ability to convey complex emotions and controlled detail content remains limited. This constraint leads to a discrepancy between the generated audio and human subjective perception in practical applications like companion robots for children and marketing bots. The core issue lies in the inconsistency between high-quality audio generation and the ultimate human subjective experience. Therefore, this challenge aims to enhance the persuasiveness and acceptability of synthesized audio, focusing on human alignment convincing and inspirational audio generation. A total of 19 teams have registered for the challenge, and the results of the competition and the competition are described in this paper.
\end{abstract}
\noindent\textbf{Index Terms}: Text-to-Speech (TTS), Audio Generation, ISCSLP 2024, ICAGC 2024, Convincing Audio

\section{Introduction}

The Inspirational and Convincing Audio Generation Challenge 2024 (ICAGC 2024) is a pioneering event in the realm of synthetic audio, set to be a highlight within the ISCSLP 2024 Competitions and Challenges track \cite{icagc2024}. Despite the significant advancements in text-to-speech (TTS) technology, current systems often fall short in conveying intricate emotions and controlled, detailed content, which is crucial for applications such as companion robots for children and marketing bots. This gap leads to a noticeable discrepancy between the generated audio and human subjective perception, particularly in real-world applications.

Previous audio synthesis competitions have laid the groundwork for advancing TTS technology, pushing the boundaries of naturalness and intelligibility. The LIMMITS 24 Challenge \cite{limmit24}, for instance, provided participants with 80 hours of TTS data across multiple languages, allowing for extensive voice cloning and multilingual synthesis experiments. Similarly, the Blizzard Challenge 2023 \cite{perrotin2023blizzard} continued its tradition of evaluating corpus-based speech synthesis systems, focusing on tasks such as building synthetic voices from provided speech data and performing extensive listening tests to evaluate their performance.

ICAGC 2024 aims to bridge this gap by challenging participants to enhance the persuasiveness and acceptability of synthesized audio. The focus is on generating audio that is not only high in quality but also capable of inspiring and resonating with listeners on a deeper emotional level. The challenge is divided into two tracks: Inspirational Emotion Modulation and Convincing Background Audio and Speech Fusion Generation.

Track 1, Inspirational Emotion Modulation, requires participants to clone the voice of a target speaker and modulate it to express convincing emotions across different themes, such as novel chapters and ancient Chinese poems. The goal is to inspire and resonate with the listener through the synthesized audio.

Track 2, Convincing Background Audio and Speech Fusion Generation, builds on the synthesized voices from Track 1 by incorporating realistic background sounds to enhance the audio's lifelike quality. Participants are tasked with adding background sounds that accurately represent real-life scenarios, thereby making the audio more convincing.

The evaluation of submissions will consider various aspects, including speaker similarity, emotion inspiration degree, and the convincing matching degree of audio and speech. By addressing these elements, ICAGC 2024 aims to push the boundaries of TTS technology, fostering innovation that aligns more closely with human subjective experiences.

This challenge provides a platform for researchers and developers to showcase their skills and contribute to the advancement of audio generation technology, ultimately leading to more effective and emotionally engaging applications.

\section{Challenge Task}
\subsection{Track 1: Inspirational Emotion Modulation}
\begin{itemize}
\item  Objective: Utilize given audio from a target speaker to clone their voice and modulate it to express proper convincing emotions for different theme (including novel chapters, ancient Chinese poems, etc.), which can inspire and resonate with the listener.

\item Materials Provided: The organizers will provide participants with ten target speakers, each with a single theme sentence.

\item Task: Competitors are required to clone the voice of the target speakers and modulate the cloned voice to express the several specified themes. For the testing set, ten paragraph-level texts will be provided. Participants must generate each text in the test set with the cloned voice, modulating proper inspirational emotion for each theme, and submit the corresponding audio files.
 \end{itemize}

\subsection{Track 2: Convincing Background Audio and Speech Fusion Generation}
\begin{itemize}

\item Objective: On the basis of the synthesized voices from Track 1, use background audio generation technology (TTA) to add realistic background sounds to make the audio more lifelike and convincing to real-life scenarios.

\item Task: Teams are tasked with submitting audio for each text in the test set, adding background sounds they believe most accurately represent real-life scenarios. These submissions will be further evaluated and scored based on their realism and naturalness.
 \end{itemize}

\section{Reference Audio Information}

The provided dataset includes audio files and their corresponding transcriptions.Each audio file follows the naming convention “itemID-speakerID-sentenceID.wav”, where “itemID” denotes the index of a complete literary work or a segment of it, “speakerID” indicates the specific speaker who produced the audio, and “sentenceID” represents the sentence index within the selected literary work.

The dataset covers a range of literary genres such as ancient poems, modern poems, novels, storytelling, and fairy tales. Some speakers have only one item, while others may have multiple items spanning various literary genres. To clone a specific speaker’s timbre, all of their audio samples can be utilized. It is important to note that the speaking styles can vary significantly across genres, even for the same speaker. Thus, the combination of “itemID-speakerID” effectively determines the speaker's timbre and speaking style in an audio file.

The transcription file, “reference.csv”, contains pairs of audio files and their corresponding text content, listed line by line. During the testing phase, another file, “test.csv”, will be provided. Similar to “reference.csv”, “test.csv” uses “itemID” to represent a complete literary work or part of it. However, in this context, an identical “itemID” appearing in both “reference.csv” and “test.csv” refers to a speaking style rather than the same literary work. The synthesis process involves generating text content based on the speaker index and the logical identifier “itemID”.

\section{Evaluation Metrics}

Evaluating audio generation systems is a complex task that requires considering various aspects. The following are the detailed definitions and evaluation methods for the proposed metrics in the Inspirational and Convincing Audio Generation Challenge 2024 (ICAGC 2024):

\subsection{Speaker Similarity}
Speaker similarity measures how closely the synthesized audio resembles the characteristics of a target speaker's speech, where the goal is to generate audio that sounds like the target speaker's timbre.

\begin{itemize}
\item Subjective: Native speakers listen to pairs of synthesized and target speaker utterances and score their speaker similarity on a rating scale.
\item Objective: Extract speaker embeddings from the synthesized and target speaker's speech, and compute distance metrics between them. Techniques like Resemblyzer\cite{wan2018generalized} and WavLM\cite{chen2022wavlm} will serve as objective assessment tools for evaluating speaker similarity.
 \end{itemize}

\subsection{Emotion Inspiration Degree}

This metric evaluates the ability of the synthesized speech to effectively convey the intended emotions in a way that resonates with and inspires the listener on a cognitive level. The goal is to generate synthesized speech that not only expresses the desired emotion accurately, but also has the power to stir corresponding feelings and thoughts within the listener, leaving them inspired, moved, or motivated.

\begin{itemize}
\item Subjective: Native speakers evaluate the inspirational impact and cognitive effect of the synthesized speech samples conveying different emotions. They will rate the degree to which the audio instilled the intended emotion within them and influenced their mindset or outlook.
 \end{itemize}

\subsection{Audio and Speech Convincing Matching Degree }
Audio enhancement techniques, such as generating sound effects, background sounds, and vocal fillers, aim to improve the overall quality and convincingness of the synthesized speech. This metric assesses the effectiveness of these techniques in enhancing the subjective experience of the listener.

\begin{itemize}
\item Subjective: Native speakers evaluate the generated non-speech audio elements, such as sound effects, background sounds, and vocal fillers, in the synthesized speech samples. They will evaluate how convincingly these audio elements blend with and complement the synthesized speech, contributing to an authentic, realistic experience as if occurring in a true-to-life setting or scenario.
 \end{itemize}

\subsection{Overall Evaluation}
The overall evaluation combines the individual metrics to provide a comprehensive assessment of the system's performance, considering trade-offs and interactions between different aspects, such as speaker similarity, emotion inspiration, and audio-speech convincing matching.

\begin{itemize}
\item Subjective: Native speakers provide holistic judgments on the overall quality of the synthesized speech.
\item Objective: Combine scores from individual metrics with weights of 0.3 for speaker similarity, 0.3 for emotion inspiration, and 0.4 for audio-speech convincing matching to obtain an overall system score.
 \end{itemize}
 
\section{Challenge Rules}

\subsection{Training Dataset}
\begin{itemize}
\item You are allowed to use external data in any way you wish. Only voice cloning reference speaker audio data will be provided by organizers.
\item Each participant \textbf{MUST} provide detailed explanations regarding the data sources and scale used, as well as other relevant details.
\end{itemize}

\subsection{Submitted Synthetic Speech Samples}
\begin{itemize}
\item Synthetic speech may be submitted at any standard sampling rate (but always at 16 bits per sample). Waveforms will not be downsampled for the listening test.
\item Any examples that you submit for evaluation will be retained by the organisers for future use.
\item You must include in your submission of the test sentences a statement of whether you give the organisers permission to publically distribute your waveforms and the corresponding listening test results in anonymised form. (We strongly encourage you to agree to this rule.)
\end{itemize}

\subsection{Paper}
\begin{itemize}
\item Each participant will be expected to submit a five-page paper (using the ISCSLP 2024 template) describing their entry for review.
\item Papers should describe the system, as well as the use of:
\begin{itemize}
\item External data, if any (e.g., other speech or text corpora)
\item Existing tools, software and models (e.g., text analysers, FastSpeech, Vall-E, ...)
\end{itemize}
\item One of the authors of each accepted paper should present it at the ISCSLP 2024.
\end{itemize}

\subsection{Listening Test}

\begin{itemize}

\item Participants who wish to submit multiple systems (e.g., an individual entry and a joint system) should contact the organisers in advance to agree with this. We will try to accommodate all reasonable requests, provided the listening test remains manageable.
\end{itemize}

\subsection{Notice}
\begin{itemize}

\item This is a challenge, which is designed to answer scientific questions, and not a competition. Therefore, we rely on your honesty in preparing your entry.
\item If you are in any doubt about how to apply these rules, please contact the organizers immediately.
\end{itemize}



\section{Challenge Submission}
\subsection{File Naming and Submission Format}
\begin{enumerate}
    \item \textbf{File Naming Convention}:      \\
    The naming format for an audio file is “type(s/p)-itemID-speakerID-sentenceID.wav”.
    \begin{itemize}
        \item \textbf{type}: Indicates whether the file is a sentence (s) or paragraph (p).
        \item \textbf{itemID}: Index of a complete literary work or a part of it.
        \item \textbf{speakerID}: Index of a specific speaker who produced that audio file.
        \item \textbf{sentenceID}: Sentence index of the selected literary work.
    \end{itemize}

    \item \textbf{Submission Requirements}
    \begin{itemize}
        \item The text to be synthesized is included in the attachment. Please ensure that you use this text for your audio synthesis. 
        \item Audio files need to be synthesized separately for sentences and paragraphs.
        \item The generated audio style must be consistent with the given reference audio. Specifically, the style of the generated audio should match the reference audio with the same itemID and speakerID.
        \item Each file should be synthesized according to the provided text and named accordingly using the format “type-itemID-speakerID-sentenceID” (e.g., s-1-1-1.wav/ p-1-1-1.wav).
        \item Place sentence files in a folder named “sentence” and paragraph files in a folder named “paragraph”.
        \item Package both folders into a single zip file. The naming format for the zip file should be “team name-track” (e.g., ICAGC-Track1). If participating in both tracks, submit two separate zip files.
    \end{itemize}
\end{enumerate}

\subsection{Submission Procedure}
Here are the steps that every participating team has to follow.

\begin{enumerate}
    \item Each team should complete the {registration-form} . The following information is mandatory:
    \begin{itemize}
        \item preferred team name - the organisers may adjust this so that all teams have meaningful, unique names
        \item affiliation - the name of your University and lab, or your Company
        \item the name of the main contact person - this must be exactly one person who is responsible for all communication with the organiser
        \item contact details: main contact person's email address, postal address, phone number
    \end{itemize}

    \item After receiving the confirmation email from the organizer, teams would receive the sample data by the given date. It should be noted that the provided data should be used only for this challenge. The data cannot be distributed to the web without the permission of the organizers.

    \item During the testing phase, we will send an email to the participating teams with 10 texts to be synthesized (including novel chapters, ancient Chinese poems, etc.). Each participating team is asked to generate the corresponding audio files using their best models.

    \item Each team has to compress those 10 generated audio files into a zip file name after the team, and then send it to the official email:
\texttt{icagc2024@iscslp2024.com}
\end{enumerate}


\begin{table*}[h!]
\caption{Track 1 Result}
\label{tab:tarck1}
\centering
\resizebox{0.9\textwidth}{!}{
\begin{tabular}{|l|c|c|c|c|c|}
\hline
Team Name    & Quality & Similarity & Emotion & MOS(avg) & MOS(std) \\ \hline
\begin{CJK}{UTF8}{gbsn}油菜花队伍\end{CJK}        & 3.89    & 3.83       & 3.85    & 3.86     & 0.22     \\ \hline
NPU-HWC          & 3.84    & 3.48       & 3.58    & 3.63     & 0.30     \\ \hline
zyzx\_ai     & 3.67    & 3.50       & 3.50    & 3.56     & 0.42     \\ \hline
PeiyangTTSer & 3.64    & 3.52       & 3.28    & 3.48     & 0.39     \\ \hline
Happy Happy  & 3.33    & 3.45       & 3.33    & 3.37     & 0.36     \\ \hline
\end{tabular}
}
\end{table*}

\begin{table*}[h!]
\caption{Track 2 Result}
\label{tab:tarck2}
\centering
\resizebox{0.9\textwidth}{!}{
\begin{tabular}{|l|c|c|c|c|c|}
\hline
Team Name    & \multicolumn{1}{l|}{Similarity} & \multicolumn{1}{l|}{Emotion} & \multicolumn{1}{l|}{Matching} & MOS(avg) & MOS(std) \\ \hline
NPU-HWC          & 4.15                            & 3.66                         & 3.33                          & 3.67     & 0.68     \\ \hline
PeiyangTTSer & 3.33                            & 3.33                         & 2.66                          & 3.06     & 0.41     \\ \hline
\end{tabular}
}
\end{table*}

\section{Results}
\subsection{Summary of submitted systems and evaluation results for Track 1}
Track 1 aims to clone a given speaker's voice and adjust the synthesized speech's emotion according to different topics. Team 1 introduces an expressive TTS synthesizer to address the Track 1 challenge. The entire framework is a cascaded system. First, the LLM model predicts speech tokens constructed by multilingual Wav2Vec and K-means. Then, a variational autoencoder (VAE) as a voice conversion model is applied to generate the mel spectrogram from the speech tokens. Finally, the HifiGAN vocoder is used to synthesize the waveform using the generated mel spectrogram as input. To encourage the disentanglement of timbre and style, their strategy is to limit the responsibilities of the autoregressive model to capturing phoneme content, duration, and prosody, while designing a separate token-to-mel decoder for reconstructing the speaker's identity. Additionally, they propose a data processing pipeline to transform in-the-wild speech data into high-quality training data with annotations for speech generation. Team 2 implements a zero-shot speaker and style cloning speech generator in Track 1, similar to Tortoise TTS, specifically designed for zero-shot speaker and style cloning. They use Single-Codec as the speech tokenizer for their proposed method, which explicitly decouples timbre and speaking style at the token level, thereby reducing the acoustic modeling burden on the autoregressive model. Additionally, they adopt DSPGAN to upsample 16k mel spectrograms to 48k high-fidelity waveforms. Team 3 mainly uses the pre-trained GPT-SoVITS, a large-scale, two-stage speech synthesis system. The system structure includes two parts: an autoregressive (AR) model and a non-autoregressive (NAR) model. The AR model focuses on extracting semantic information from the synthesized speech, such as content and style, while the NAR model focuses on controlling non-linguistic elements in the synthesized output, such as timbre and emotion. They propose a few-shot learning strategy for specific topics, fine-tuning the pre-trained model with sentences from different speakers on the same topic to optimize the model for particular themes and individual speaker characteristics. Team 4 proposes Hola-TTS, a robust speech synthesis model for cross-lingual zero-shot speech synthesis, suitable for four languages: Chinese, English, Japanese, and Korean. Hola-TTS uses a training approach based on large language models (LLMs). It first discretizes the speech signal into discrete tokens using the EnCodec encoder. Simultaneously, text input is converted into a sequence of phonemes with phoneme-level language labels. Both are concatenated and fed into a Transformer decoder for predicting the next token. During inference, tokens are predicted in an autoregressive manner, and the EnCodec decoder reconstructs the audio. Specifically, while large models are not particularly sensitive to data quality, clean data still improves the quality of the generated output. Therefore, they perform data cleaning before training. The result of Track1 are listed in Table.~\ref{tab:tarck1}.

\subsection{Summary of submitted systems and evaluation resultsfor Track 2}
Track 2 aims to add realistic background sounds to the speech, making the audio more lifelike and convincing in real-life scenarios. Team 5 develops a large language model (LLM)-based cascading system to generate appropriate background audio for speech synthesis, consisting mainly of two stages: text-to-description and description-to-audio. In the text-to-description stage, the goal is to use the LLM's semantic understanding to generate descriptive text corresponding to the speech transcript. For this, they select DeepSeek-V22, a robust expert mixture model (MoE) known for its strong language understanding capabilities. In the description-to-audio stage, they use Tango 2 to generate the background audio. Additionally, since the main focus of the audio is the speech content, they apply a 10dB gain reduction to the generated background audio during the mixing process to ensure it does not overshadow the primary speech. Team 6 builds an innovative system that deeply integrates the GPT-SoVITS1 and AudioLDM 2 models to achieve highly emotional and realistic background audio for text-to-speech synthesis. They innovatively combine GPT-3.5 with AudioLDM 2 in Track 2. Based on the results achieved from fine-tuning GPT-SoVITS, the system uses LLM’s deep understanding of text content to intelligently match each segment of speech with the most appropriate background description. Then, AudioLDM 2 generates these background sounds, ensuring they perfectly blend with the speech in terms of style, rhythm, and other aspects. The result of Track2 are listed in Table.~\ref{tab:tarck2}.

\section{Conclusion}
The ICAGC 2024 Challenge successfully advances the development of synthetic audio technology, particularly in enhancing the emotional conveyance and persuasiveness of audio. Through the two tracks, we observe the application of various innovative approaches that not only improve the quality of synthetic audio but also enhance emotional resonance with listeners. The diversity and innovation of the entries indicate that TTS technology is evolving towards a more humanized and personalized direction. The results of the challenge highlight the important role of evaluation metrics in driving technological development. 

\section{Acknowledgements}

This work is supported by the National Natural Science Foundation of China (NSFC) (No.62101553, No.62306316, No.U21B20210, No. 62201571).
\vfill
\newpage

\bibliography{mybib}


\end{document}